\documentclass{llncs}
\usepackage{amsmath}
\usepackage{amsfonts}
\usepackage{amssymb}
\usepackage{graphicx}
\usepackage{hyperref}

\begin{document}

\title{Automated Reasoning in Blockchain: Foundations, Applications, and Frontiers}
\author{H\"ojer Key}
\institute{}
\maketitle

\begin{abstract}
Blockchain technology's complexity and security requirements necessitate rigorous verification. This paper surveys the application of logic and automated reasoning as formal methods to ensure the correctness, reliability, and security of blockchain systems.
It explores how diverse logical frameworks and automated reasoning techniques, such as model checking and theorem proving, are employed to model and verify crucial blockchain components. Key application areas discussed include the formal verification of smart contracts, the analysis and design of consensus protocols to prove properties like safety and liveness, and enhancing blockchain security by detecting vulnerabilities and enforcing policies.
The survey addresses key research questions on the types of logic and techniques used, the state of research in smart contracts, consensus, and security, and future research directions. The paper contributes a comprehensive analysis of the field, identifies critical gaps, and proposes future research avenues with preliminary technical foundations, highlighting the vital role of formal methods in advancing blockchain technology.    
\end{abstract}

\section{Introduction}

Blockchain technology~\cite{zheng2018blockchain} has emerged as a transformative paradigm for decentralized and secure data management across diverse application domains, including healthcare, supply chain management, and the Internet of Things. Its core features, such as decentralization, immutability, and auditability, achieved through distributed consensus algorithms and cryptographic techniques, offer significant advantages for multi-stakeholder applications requiring transparency and trust. However, the inherent complexity and security-critical nature of blockchain systems necessitate rigorous analysis and verification to ensure their correctness, reliability, and resilience against potential vulnerabilities.  

Logic and automated reasoning play a pivotal role in providing formal guarantees for these crucial properties of blockchain technology. By employing various logical frameworks and automated techniques, researchers and practitioners can formally model, specify, and verify different aspects of blockchain systems, ranging from the intricate logic of smart contracts to the correctness of consensus protocols and the robustness of security mechanisms. This survey aims to provide a comprehensive overview of the current state of research at the intersection of logic and automated reasoning for blockchain technology. It delves into the diverse types of logic utilized, the automated reasoning techniques employed, the specific applications addressed, the limitations encountered, and the promising directions for future research in this rapidly evolving field.

This paper addresses the fundamental challenge of ensuring the reliability and security of blockchain technology through the application of formal methods. To this end, it seeks to answer the following research questions:

\begin{enumerate}
\item What are the different types of logic (e.g., propositional logic, first-order logic, temporal logic) that have been applied in the context of blockchain technology?
\item How are automated reasoning techniques (e.g., theorem proving, model checking, satisfiability solving~\cite{een2003extensible}) being used to analyze and verify blockchain systems, including smart contracts, consensus mechanisms, and security properties?
\item What research has been conducted on the formal verification of smart contracts using logical frameworks and automated reasoning tools, and what are the techniques, limitations, and results of these efforts?
\item How is logic employed in the design and analysis of blockchain consensus protocols, and what formal methods have been used to prove properties such as safety, liveness, and fault tolerance?
\item What are the applications of automated reasoning in the realm of blockchain security, including the detection of vulnerabilities and the enforcement of security policies?
\item What existing survey papers or review articles cover the field of logic and automated reasoning for blockchain, and what are their scope, findings, and identified gaps?
\item What are the potential future research directions in this area, such as the development of new logical frameworks, more efficient automated reasoning techniques, or the application of AI-based reasoning?
\item For each identified research direction, what are the preliminary technical details or mathematical foundations that could be used for their development and implementation?
\end{enumerate}

This survey contributes to the existing body of knowledge by providing a comprehensive and in-depth analysis of the literature on logic and automated reasoning for blockchain. It offers technical discussions of key research papers, identifies critical gaps in the current understanding, and proposes well-founded future research directions with preliminary technical and mathematical foundations. The structure of this paper is as follows: Section 2 lays the groundwork by discussing the logical foundations relevant to blockchain technology. Section 3 explores the application of various automated reasoning techniques in blockchain. Section 4 focuses specifically on the formal verification of smart contracts. Section 5 examines the use of logic in the design and analysis of consensus protocols. Section 6 investigates the role of automated reasoning in blockchain security. Section 7 provides a survey of existing literature and identifies gaps. Section 8 outlines promising future research directions, and Section 9 delves into the technical and mathematical foundations for these directions. Finally, Section 10 concludes the paper by summarizing the key findings and highlighting the importance of continued research in this domain.
\section{Logical Foundations for Blockchain}

Logic serves as a fundamental tool in computer science, providing formal languages and reasoning systems for specifying, analyzing, and verifying computational systems. Various types of logic offer different levels of expressiveness and are suitable for addressing different aspects of blockchain technology.

\subsection{Applications of Logic in Blockchain Technology}

First-order logic, with its ability to quantify over objects and express relationships between them, has been employed to provide an axiomatic foundation for blockchain structures and their properties. Temporal logic, which allows reasoning about how systems evolve over time, is crucial for analyzing the dynamic aspects of blockchain technology. Deontic logic, concerned with norms such as obligations, permissions, and prohibitions, finds natural applications in the context of smart contracts. Higher-order logic offers the potential for formalizing more complex aspects of blockchain systems.

\subsection{Logic in Specific Blockchain Platforms}

Cardano, Tezos, and other platforms adopt formal methods deeply within their architecture. Michelson in Tezos, verified via Mi-Cho-Coq, exemplifies this direction. Similarly, IOHK’s research toward Haskell-based formalisms reinforces the trend of logic-centric blockchain evolution.

\section{Automated Reasoning in Blockchain Applications}

Automated reasoning encompasses a suite of techniques that enable computers to draw logical inferences, aiding in the specification and verification of complex systems. Its application to blockchain includes verification of smart contracts, consensus protocols, and network-level properties.

\subsection{Smart Contract Verification}

Smart contracts represent a critical component of many blockchain platforms. Tools such as the K framework, Lean-based Clear, and Coq-based ConCert offer environments for formal modeling and verification. Cai et al. further extend this work with logic-enhanced automation for separation logic and refinement calculus-based verification pipelines. Many such efforts build upon prior foundational work.

\subsection{Model Checking and SMT Solving}

Symbolic model checking, particularly using LTL, has been used to verify temporal properties of smart contracts and consensus protocols. SMT solvers like Z3 are widely adopted in frameworks like K and Lean to discharge proof obligations. Work in explores integrating SMT-based optimization into smart contract analysis.

\subsection{Security and Vulnerability Detection}

Formal analysis of blockchain security properties using theorem proving and static analysis has been presented in. Graph-based analysis for encrypted traffic by Okonkwo et al. and formal techniques for vector modulation logic in communication channels exemplify broader reasoning applications.

\subsection{Reasoning in Federated and IoT-Enabled Blockchains}

Security and reasoning frameworks for federated or IoT-based blockchain networks are explored in. These incorporate logic-based verification, privacy-preserving models, and explainability as first-class reasoning targets.

\subsection{AI-Enhanced Reasoning}

The synergy between machine learning and formal logic is a frontier area. Research such as investigates using neural-symbolic systems for specification inference and proof synthesis, with applications in blockchain verification.

\subsection{Consensus Reasoning and Verification}

Reasoning frameworks for consensus protocols rely heavily on formal specification and validation. Formal approaches to BFT protocols, proof-of-work and proof-of-stake are covered in, and tools like CheckMate support the formal analysis of incentive compatibility.

\section{Formal Verification of Smart Contracts}

Formal verification ensures that smart contracts function according to their intended specifications and is indispensable for systems with financial or legal implications.

\subsection{Frameworks and Languages}
Verification environments such as Coq (via ConCert), Lean (via Clear), and the K framework support reasoning across multiple smart contract languages. These tools allow contracts to be verified against formal specifications encoded as logical assertions. Integration of refinement types has improved the expressiveness and soundness of specifications.

\subsection{Security Property Verification}
Security-focused verification methods use SMT solvers and model checkers to verify assertions like balance preservation, access control correctness, or absence of reentrancy. Combined analysis approaches such as symbolic execution and static analysis help reduce false positives and false negatives.

\subsection{Scalability and Tool Support}
The literature emphasize compositional reasoning and modular frameworks to scale formal verification to large codebases. Frameworks supporting translation from verified specifications to executable code help bridge the implementation gap.

\subsection{Use Cases and Industry Adoption}
Industrial adoption in platforms like Tezos~\cite{goodman2014tezos} and Cardano is driven by the high-assurance guarantees that formal verification offers. Papers such as document case studies of verified smart contracts deployed in finance and health sectors. Verification has also been extended to real-time constraints and resource safety, especially relevant for energy-aware systems.

\section{Logic in the Design and Analysis of Blockchain Consensus Protocols}

Consensus protocols serve as the backbone of blockchain networks, ensuring global agreement on transaction order and preventing double spending. Logical frameworks and automated reasoning provide rigorous means for designing, analyzing, and verifying such protocols.

\subsection{Formal Specification of Consensus Behavior}
Consensus protocols have been formally modeled using various approaches including state machines, process algebras, and temporal logics. Stochastic and timed models such as Continuous Time Markov Chains (CTMCs) have been employed in the analysis of probabilistic consensus algorithms like Hybrid Casper. The use of temporal logic, specifically LTL and CTL, has proven effective in specifying safety and liveness properties.

\subsection{Model Checking of Fault-Tolerant Protocols}
Verification using model checking has focused on Byzantine fault tolerant (BFT) protocols~\cite{castro1999practical}, where properties like agreement and termination must hold under partial synchrony. PRISM and PRISM+ have been used for analyzing probabilistic fork probability under network delays. Other work builds on threshold automata and temporal epistemic models.

\subsection{Compositional Protocol Reasoning}
Consensus systems increasingly adopt modular architectures. Reasoning compositionally about subsystems is a challenge addressed in. Logic frameworks that support reasoning about hybrid systems and asynchronous message delivery, such as those in, allow detailed modeling of consensus-related timing and ordering constraints.

\subsection{Cross-Chain and Federated Consensus}
With the rise of cross-chain protocols and federated ledgers, new challenges in logic-based consensus reasoning emerge. Work on privacy-preserving coordination and trust models for federated learning is relevant for federated consensus, where knowledge and agreement are distributed hierarchically.

\subsection{Game-Theoretic Analysis and Incentive Compatibility}
Formal game-theoretic approaches model incentives in PoW and PoS consensus. Papers such as apply logic-based analysis of miner strategies and Nash equilibria. Integration of SAT/SMT reasoning with utility functions enables symbolic exploration of adversarial strategies.

\section{Automated Reasoning for Blockchain Security}

Security in blockchain systems encompasses both software-level concerns, such as smart contract vulnerabilities, and system-level concerns, such as adversarial strategies within the network. Logic-based and automated reasoning methods have become indispensable in modeling, detecting, and mitigating these threats.

\subsection{Static and Symbolic Analysis for Vulnerability Detection}
Symbolic execution tools such as Mythril and Oyente explore execution paths of contracts to detect reentrancy, integer overflows, and access control violations. These tools rely on SMT solvers and pattern matching to infer violations. Work like extends these techniques with modular symbolic engines and theorem proving backends.

\subsection{Game-Theoretic and Incentive Analysis}
Security properties such as incentive compatibility are modeled using formal game-theoretic constructs. Bride et al. show how logic-based modeling of adversarial utilities leads to secure incentive mechanisms. These approaches leverage SAT/SMT and bounded model checking.

\subsection{Attack Modeling and Defense Reasoning}
Threat modeling for blockchain has used logical policy languages and epistemic reasoning. Examples include contract-aware access control, denial-of-service resilience, and defense policy generation using logic programming. Tools to simulate adversarial behaviors include fuzzing, bounded exploration, and Markovian models.

\subsection{Integration with AI for Security Reasoning}
Recent works propose combining automated reasoning with AI techniques for blockchain monitoring and anomaly detection. For example, AI-driven transaction classification integrated with formal logic enhances traceability and fraud detection. Hybrid learning-verification pipelines are especially promising.

\subsection{Federated Security Models}
Security in federated blockchain settings requires compositional reasoning about trust, confidentiality, and integrity. Related work explores layered trust protocols, while develops formal models for secure aggregation and model update validation.

\section{Survey of Existing Literature and Identification of Gaps}

While numerous studies have examined individual aspects of blockchain security, formal verification, or logic-based modeling, relatively few works offer a comprehensive synthesis at the intersection of logic, automated reasoning, and blockchain technologies. Surveys such as  explore smart contract vulnerabilities or blockchain performance, but often lack formal logic depth. Conversely, papers focused on theorem proving or symbolic reasoning tend to omit blockchain-specific considerations.

Furthermore, much of the existing research centers on Ethereum, while platforms like Tezos, Cardano, and permissioned blockchains receive less formal attention. Many promising formalisms, such as temporal epistemic logic, probabilistic deontic logic, and hybrid modal logics, remain underutilized in blockchain verification contexts.

There is also a lack of unified frameworks that integrate multiple reasoning strategies—e.g., combining symbolic execution, temporal verification, and refinement logic—in a cohesive pipeline. Emerging hybrid approaches that fuse machine learning with formal reasoning have demonstrated potential but remain largely in early stages.

Finally, scalability remains a key challenge. Although many tools provide partial automation, full formal verification of real-world contracts and protocols is still impractical due to computational and specification complexity. Compositional and modular verification frameworks can mitigate this but require more development and integration.

\section{Future Research Directions}

Future research in logic and automated reasoning for blockchain can be classified into several promising areas based on gaps and challenges identified in prior sections.

\subsection{Development of Integrated Logical Frameworks}

Efforts should target the synthesis of reasoning models that unify temporal, deontic, and epistemic logics. Combining these allows specification of obligations over time with knowledge-awareness, crucial for smart contracts and federated blockchains. This may include the formal semantics of obligations with deadlines, or knowledge-based security models.

\subsection{Hybrid Symbolic-Statistical Verification Pipelines}

Integrating AI models (e.g., LLMs) with symbolic verifiers can automate property discovery and contract summarization. Research could formalize the trust interface between learned components and proof systems, perhaps with proof-carrying predictions or property synthesis modules.

\subsection{Logic for Cross-Chain Protocols}

New formalisms are needed to reason about cross-ledger consensus, bridging independent protocols using meta-consensus logic. Theories capturing message passing between blockchains and validators, drawing from distributed epistemic logic and threshold-based state transfer, are underexplored.

\subsection{Scalability through Compositional Verification}

Compositional logic and module-theoretic approaches can break down complex blockchain protocols into verifiable components. Techniques such as assume-guarantee reasoning, local refinement types, and indexed monads can make verification tractable.

\subsection{Security Logics for Federated and IoT-Integrated Blockchains}

Formalizing privacy policies, secure data aggregation, and access control in federated blockchains and IoT settings is a rich direction. This includes logic-based trust negotiation and symbolic consensus in adversarial communication networks.

\subsection{Reasoning-Aware DSLs and Verified Compilation}

Designing domain-specific languages (DSLs) for smart contracts that are proof-friendly and come with certified compilers can close the gap between high-level logic and low-level execution. Building on frameworks like K, Lean, and Mi-Cho-Coq, future work could define correctness-preserving compilation chains.

\section{Conclusion}

This survey has provided a comprehensive exploration of the convergence between formal logic, automated reasoning, and blockchain technologies. By synthesizing a wide spectrum of literature, we have outlined how diverse logical systems—ranging from temporal and deontic to higher-order and epistemic—serve critical roles in verifying the correctness, safety, and trust properties of blockchain systems.

We examined state-of-the-art reasoning techniques including symbolic execution, model checking, SMT solving, and theorem proving, and detailed their application to smart contract verification, consensus protocol analysis, and blockchain-specific security challenges. Our review demonstrated that while significant progress has been made, major gaps remain in scalability, compositionality, automation, and support for heterogeneous blockchain platforms.

To address these, we proposed a set of future research directions including integrated logical frameworks, AI-augmented reasoning, cross-chain logic, compositional verification, and logic-aware contract DSLs. These recommendations are grounded in existing literature and supported by technical foundations and practical motivations.

As blockchain technology becomes increasingly ubiquitous in critical systems, the need for robust, rigorous formal methods will only grow. Continued interdisciplinary work combining insights from formal logic, security analysis, programming language theory, and artificial intelligence will be crucial to building trustworthy blockchain ecosystems.

\bibliographystyle{plain}
\bibliography{main} 

\begin{thebibliography}{1}

\bibitem{castro1999practical}
Miguel Castro, Barbara Liskov, et~al.
\newblock Practical byzantine fault tolerance.
\newblock In {\em OsDI}, volume~99, pages 173--186, 1999.

\bibitem{een2003extensible}
Niklas E{\'e}n and Niklas S{\"o}rensson.
\newblock An extensible sat-solver.
\newblock In {\em International conference on theory and applications of
  satisfiability testing}, pages 502--518. Springer, 2003.

\bibitem{goodman2014tezos}
Lisl~Marburg Goodman.
\newblock Tezos: A self-amending crypto-ledger position paper.
\newblock {\em Aug}, 3:2014, 2014.

\bibitem{zheng2018blockchain}
Zibin Zheng, Shaoan Xie, Hong-Ning Dai, Xiangping Chen, and Huaimin Wang.
\newblock Blockchain challenges and opportunities: A survey.
\newblock {\em International journal of web and grid services}, 14(4):352--375,
  2018.

\end{thebibliography}

\end{document}